\documentclass [6pt,a4paper]{article}
\usepackage{bbm}
\usepackage{amsfonts}
\usepackage{mathrsfs}
\usepackage {amssymb}
\usepackage {amsmath}
\usepackage{amsthm}
\usepackage{latexsym}
\def\dse#1{\vskip 0.6cm\noindent
        {\large\bf #1}
        \vskip 0.4cm}
\def\dse#1{\vskip 0.6cm\noindent
        {\large\bf #1}
        \vskip 0.4cm}
\usepackage{lastpage}
\usepackage{epsfig}

\begin{document}
\begin{center}
\textbf{\large{The reversible negacyclic codes
over finite fields}}\footnote {E-mail
addresses:
 zhushixin@hfut.edu.cn(S.Zhu), pbbmath@126.com(B.Pang), sunzhonghuas@163.com(Z. Sun).
 This research is supported by the National Natural Science
Foundation of China (No.61370089;\ No.61572168).}\\
\end{center}

\begin{center}
{ Shixin Zhu, Binbin Pang,  Zhonghua Sun}
\end{center}

\begin{center}
\textit{Department of Mathematics, Hefei University of
Technology, Hefei 230009, Anhui, P.R.China }
\end{center}

\noindent\textbf{Abstract:} In this paper, by investigating the factor of the $x^n+1$, we deduce that the structure of the reversible negacyclic code over the finite field $\mathbb{F}_{q}$, where $q$ is an odd prime power. Though studying $q-$cyclotomic cosets modulo $2n$, we obtain the parameters of negacyclic BCH code of length $n=\frac{q^\ell+1}{2}$ , $n=\frac{q^m-1}{2(q-1)}$ and $n=\frac{q^{t\cdot2^\tau}-1}{2(q^t+1)}$. Some optimal linear codes from negacyclic codes are given. Finally, we discuss a class of MDS LCD negacyclic codes.\\

\noindent\emph{Keywords}: Negacyclic codes,  Reversible code, Negacyclic BCH codes, MDS code.
\dse{1~~Introduction}
The negacyclic codes have been well studied in literatures. And the definition of negacyclic BCH code was given in [4]. Dinh established the structure of negacyclic codes of length $n$, where gcd$(n,q)=1$ [11]. Aydin, Sliap and Ray-chaudhuri gave the BCH bound for the constacyclic codes [3]. LCD codes were initiated by Massey [6], he also showed the existence of the asymptotically good LCD codes. The condition of the LCD codes was given by Yang and Massey [7]. Hou and Oggier acquired the construction and properties of a lattice from LCD codes [9]. Lina and Nocon constructed some special LCD codes and confirmed that permutation equivalence of codes preserves the LCD-ness of codes. Ding and Li constructed several classes of reversible cyclic codes over finite fields and analyzed their parameters [1], and they also showed the parameters of some reversible BCH codes [5]. G\"{u}neri and \"{O}zkaya studied the quasi-cyclic complementary dual code by using their concatenated structure, and they also constructed quasi-cyclic complementary dual code from codes over larger alphabets [13]. The existence of the MDS Hermitian self-orthogonal and self-dual was obtained by Yang and Cai [12].

We will study the reversible negacyclic codes over finite fields. In this paper, using the method of investigate LCD cyclic codes, we deduce the condition of reversible negacycylic codes. The structure of LCD negacyclic codes is determined, and in the special case, the quantity of reversible negacyclic codes is gained. We discuss the parameters of negacyclic BCH codes when the length $n=\frac{q^\ell+1}{2}$, $n=\frac{q^m-1}{2(q-1)}$ and $n=\frac{q^{t\cdot2^\tau}-1}{2(q^t+1)}$, and a class of MDS LCD negacyclic codes.

\dse{2~~Preparation}
Throughout this paper, let
$\mathbb{F}_q$ be a finite field of size $q$,  where $q$ is an odd prime power. A linear $[n,k,d]$ code $C$ over $\mathbb{F}_q$ is called negacyclic if $(c_0,c_1,\ldots,c_{n-1})\in C$ implies its negacyclic shift
$(-c_{n-1},c_0,\ldots,c_{n-2})\in C$. Let $C$ be an $[n,k]$ linear code over $\mathbb{F}_q$, its dual code $C^\perp$ is defined by $C^{\perp}=\{{\textbf{u}\in \mathbb{F}_q^n\mid \textbf{u}\cdot \textbf{c}=\textbf{0},\forall\  \textbf{c}\in C}\}$. By identifying any vector $(c_0,c_1,\ldots,c_{n-1})\in \mathbb{F}_q$ corresponds to a polynomial $c_0+c_1x+,\cdots,+c_{n-1}x^{n-1}\in \mathbb{F}_q[x]/\langle x^n+1\rangle$, then a linear negacyclic code over $\mathbb{F}_q$ is an ideal of ring $\mathbb{F}_q[x]/\langle x^n+1\rangle$.
In fact every ideal in  $\mathbb{F}_q[x]/\langle x^n+1\rangle$ is a principal ideal, so every negacyclic code $C$ has generator polynomial $g(x)$. Let $C=\langle g(x)\rangle$, where $g(x)$ is a unique monic and has minimal degree polynomial in $C$. And $h(x)=(x^n+1)/g(x)$ is referred to as the check polynomial of $C$. The dual code of $C$ is also negacyclic code and has generator polynomial $g^\perp(x)=x^{\textrm{deg}h}h(x^{-1})$.

In this paper, we always assume that gcd$(n,q)=1$, and note that $x^n+1$ has no repeated root over $\mathbb{F}_q$ if and only if gcd$(n,q)=1$. Let $\gamma$ be a primitive $2n-$th root of unity in $\mathbb{F}_{q^m}$, where $m$ is the multiplicative of $q$ modulo $2n$, i.e. $m=\textrm{ord}_{2n}(q)$. Then the roots of $x^n+1$ are $\gamma^{1+2i}$,  $0\leq i\leq n-1$. Let $\mathbb{Z}_{2n}=\{0,1,\cdots,2n-1\}$. For any $s\in \mathbb{Z}_{2n}$, the $q-$cyclotomic coset$$C_s=\{s,sq,sq^2,\cdots,sq^{d_s-1}\}\ \textrm{mod}2n\subseteq \mathbb{Z}_{2n}$$
where $d_s$ is the smallest positive integer such that $sq^{d_s}\equiv s\ \textrm{mod}2n$, and the size of the $C_s$. Let $T=\{1+2i\mid0\leq i\leq n-1\}$, containing all the odd integers of $\mathbb{Z}_{2n}$, obviously, $T\subseteq \mathbb{Z}_{2n}$ and $|T|=\frac{1}{2}|\mathbb{Z}_{2n}|$. \\

\noindent\textbf{Lemma 2.1.} For any $s\in \mathbb{Z}_{2n}$, $T\cap C_s=C_s\ \textrm{or}\  \ \emptyset$, where $C_s$ denotes the $q-$cyclotomic coset modulo $2n$.\\

Let $T_s=T\cap C_s$, if $T\cap C_s=C_s$. Let $X_{(n,q)}$ be a set of all the coset leaders of $T_s$, we have any $s,t\in X_{(n,q)}, s\neq t$, $T_s\cap T_t=\emptyset$, and
\begin{equation}
\large{\bigcup_{s\in X_{(n,q)}}T_s=T}\tag{2.1}
\end{equation}
\noindent\textbf{Lemma 2.2.} The cardinality $d_s$ of $T_s$ is a divisor of $m=\textrm{ord}_{2n}(q)$ which is equivalent to $d_1=|T_1|$.\\

\noindent\textbf{Theorem 2.3.} Let $n$ be an odd integer such that gcd$(n,q)=1$. For any $s\in X_{(n,q)}$, then $|T_s|=|C_{2s}|$. \\

\noindent\textbf{Proof}.\ From the definition of $T_s$, $T_s=C_s=\{s,sq,sq^2,\cdots,sq^{d_s-1}\}\ \textrm{mod}2n$, where $d_s$ is the smallest positive integer such that $sq^{d_s}\equiv s\ \textrm{mod}2n$. We easily obtain $2sq^{d_s}\equiv 2s\  \textrm{mod}2n$. where $d_s$ must also be the smallest positive integer.  Otherwise, there exist an integer $d'_s$, and $d'_s<d_s$, such that $2sq^{d'_s}\equiv 2s\ \textrm{mod2}n$, we get that there exist an integer $k$ such that $2sq^{d'_s}-2s=2nk$. Since $s,q,n$ are all odd integers, we deduce $4|(2sq^{d'_s}-2s)$, then $4|2nk$. Thus $sq^{d'_s}\equiv s\ \textrm{mod}2n$, which is contrary to the definition of $d_s$, the conclusion is obtained.\qed\\

The following is immediate from [1] Lemma 2.\\

\noindent\textbf{Lemma 2.4.} Let $q^{\lfloor m/2\rfloor}/2<n\leq (q^m-1)/2$ be a positive integer such that gcd$(n,q)=1$, where $m=\textrm{ord}_{2n}(q)$. Then the cardinality of $T_s=T\cap C_s=\{sq^i\ \textrm{mod}2n|0\leq i\leq m-1\}$ is equal to $m$ for $\forall s\in X_{(n,q)}$ in the range $1\leq s\leq 2nq^{\lfloor m/2\rfloor}/(q^m-1)$. In this case every $s$ with $s\not\equiv 0\ \textrm{mod}q$ is a coset leader of $T_s$.\\

Let $\alpha$ be a generator of $\mathbb{F}_{q^m}^\ast$, where $m=\textrm{ord}_{2n}(q)$. Put $\beta=\alpha^{(q^m-1)/2n}$, then $\beta$ is a primitive $2n-$th root of unity in $\mathbb{F}_{q^m}$, the minimal polynomial $m_i(x)$ of $\beta^i, i\in X_{(n,q)}$ over $\mathbb{F}_q$ is given by $m_i(x)=\Pi_{j\in T_i}(x-\beta^j)$. Summarizing the equality (2.1) gets
\begin{equation}
x^n+1=\prod_{i\in X_{(n,q)}}m_i(x)\tag{2.2}
\end{equation}
which is the canonical factorization of $x^n+1$ over $\mathbb{F}_q$. This is vital for studying of negacyclic codes. A linear code has complementary dual (or LCD code for short) if Hull$(C)=C\cap C^\perp=\{\textbf{0}\}$, which is equivalent to $C+C^\perp=\mathbb{F}_q^n$.\\

\dse{3~~The structure of LCD negacyclic code }  Let $h(x)=a_nx^n+a_{n-1}x^{n-1}+\cdots+a_0\in \mathbb{F}_q[x]$, with $a_n\neq 0$ and $ a_0\neq 0$, the reciprocal polynomial $h^\ast(x)$ of $h(x)$  is defined by $$h^\ast(x)=a_0^{-1}x^nh(x^{-1}).$$
\noindent\textbf{Lemma 3.1.}  Let $h(x), f(x)\in \mathbb{F}_q[x]$. Then

1) If $\textrm{deg}h\geq \textrm{deg}f$, then $(h(x)+f(x))^\ast=h^\ast(x)+x^{\textrm{deg}h-\textrm{deg}f}f^\ast(x)$,

2) $(h(x)f(x))^\ast=h^\ast(x)f^\ast(x)$.\\

A code $C$ is called reversible if $(c_0,c_1,\ldots,c_{n-1})\in C$ implies that $(c_{n-1},c_{n-2},\ldots,c_0)\in C$. Note that a negacyclic code $C$ is reversible if and only if the generator polynomial of $C$ is self-reciprocal.\\

\noindent\textbf{Theorem 3.2.}  Let $C$ be a negacyclic code over $\mathbb{F}_q $ with generator polynomial $g(x)$, then the following conclusions are equivalent.

1) $C$ is a LCD code,

2) $g(x)$ is self-reciprocal $(g(x)=g^\ast(x))$,

3) An element $\beta$ in the splitting field of $g(x)$, if $g(\beta)=0$, then $g(\beta^{-1})=0$.\\

\noindent\textbf{Proof}.\ 1) is equivalent to $C+C^\perp=\mathbb{F}_q^n$, if and only if $C=\langle g(x)\rangle$ and $C^\perp=\langle h(x)\rangle$, where $h(x)=(x^n+1)/g(x)$. We get that $C$  and  $C^\perp$ are both reversible. It is equivalent to 2) and 3).\qed\\

\noindent\textbf{Theorem 3.3.}  The negacyclic code over $\mathbb{F}_q$ of length $n$ is reversible, if $-1$ is a power of $q$ \textrm{mod}$2n$.\\

\noindent\textbf{Proof}.\ By the definition of $T_s$, we get that there exist $\ell$ such that $sq^\ell\equiv -s\ \textrm{mod}2n$, thus $-s\in T_s$. Hence every irreducible factor of $x^n+1$ is self-reciprocal, the corresponding negacyclic code is reversible.\qed\\

\noindent\textbf{Theorem 3.4.}  The irreducible polynomial $m_s(x)$ is self-reciprocal if only and if $2n-s\in T_s$.\\

Let $Y_{(n,q)}$ be a set such that $\{T_s\cup T_{2n-s}|s\in Y_{(n,q)}\}$ is a partition of $T$. Notice that there are different choices for $Y_{(n,q)}$.\\

\noindent\textbf{Theorem 3.5.} There are $2^{|Y_{(n,q)}|}-1$ reversible negacyclic codes over $\mathbb{F}_q$ of length $n$, and those generator polynomials are as follows $$g(x)=\prod_{s\in S}\textrm{lcm}(m_s(x),m_{2n-s}(x))$$
which $S\subseteq Y_{(n,q)}$ and $S\neq \emptyset$.\\

\noindent\textbf{Example 3.6.} Let $n=7,q=3$, from the definition of $C_s$ we can deduce  $$C_0=\{0\}\ \ \ C_1=\{1,3,5,9,11,13\}$$ $$C_7=\{7\}\ \ \ C_2=\{2,4,6,8,10,12\}$$\\
Notice that $T=\{1,3,5,7,9,11,13\}$, $T_1=T\cap C_1=C_1$, $T_7=T\cap C_7=C_7$, from the equality $(2.2)$, we get that $x^7+1=m_1(x)m_7(x)$, where $m_1(x)=x^6+2x^5+x^4+2x^3+x^2+2x+1$, $m_7(x)=x+1$. In this case, $m_1(x)$ and $m_7(x)$ are both self-reciprocal. And we obtain  $X_{(n,q)}=Y_{(n,q)}=\{1,7\}$. Thus, the number of the reversible ternary negacyclic codes of length 7 is 3.\\

\noindent\textbf{Lemma 3.7.} Let $m\geq1$ and $b>1, b\in \mathbb{Z}$ . Then
\begin{eqnarray*}
\textrm{gcd}(b^n+1,b^m-1)=
\left\{ {{\begin{array}{ll}
 {1}, & {\textrm{if}\ \frac{m}{\textrm{gcd}(n,m)}}\ \textrm{is odd and}\ b \textrm{ is even},\\
 {2}, & {\textrm{if}\ \frac{m}{\textrm{gcd}(n,m)}}\ \textrm{is odd and}\ b \textrm{ is odd}, \\
 {b^{\textrm{gcd}(n,m)}+1}, & {\textrm{if}\ \frac{m}{\textrm{gcd}(n,m)}}\ \textrm{is even}. \\
\end{array} }} \right .
\end{eqnarray*}
\noindent\textbf{Theorem 3.8.} Let $q=p^\ell$, where $p$ is an odd prime integer. Put $n=\frac{(q^m-1)}{2}$ be an odd integer. Then

1) If $m$ is odd integer, $x+1$ is the only self-reciprocal irreducible factor of $x^n+1$ over $\mathbb{F}_q$,

2) If $m$ is odd prime, there are $2^{\frac{q^m+(m-1)q+m}{4m}}-1$ reversible negacyclic codes of length $n$ over $\mathbb{F}_q$.\\

\noindent\textbf{Proof}.\ 1). From Lemma 3.7, we get any $0\leq i\leq m-1$, gcd$(q^i+1,q^m-1)=2$, thus $s(1+q^i)\equiv 0\ \textrm{mod}2n$ if and only if $s=0$ or $s=n$, however, $n\in T$ and $0\not\in T$.  Hence, $x+1$ is the only self-reciprocal irreducible factor of $x^n+1$ over $\mathbb{F}_q$.\\
2). Since $m$ is odd prime, the cardinality of $T_s$ is either 1 or $m$. Since gcd$(q-1,q^m-1)=q-1$, there are exactly $(q-1)$ cycloyomic cosets modulo $2n$ have size 1. By Theorem 2.3, any $s\in X_{(n,q)}$, $|T_s|=|C_{2s}|$. We then deduce that the number of $|T_s|=1$ is $\frac{q-1}{2}$, and the number of $|T_s|=m$ is half of the number of $|C_a|=m$, where $a\in \mathbb{Z}_{2n}$. Summarizing the 1) we get  $$|Y_{(n,q)}|=\frac{\frac{q^m-1}{2}-\frac{q-1}{2}}{2m}+\frac{\frac{q-1}{2}-1}{2}+1=\frac{q^m+(m-1)q+m}{4m}$$ Thus, there are $2^{\frac{q^m+(m-1)q+m}{4m}}-1$ reversible negacyclic codes of length $n=\frac{q^m-1}{2}$ over $\mathbb{F}_q$\qed\\

Let $n$ be a positive integer such that gcd$(n,q)=1$, and we know $x^n+1=\Pi_{s\in X_{(n,q)}}m_s(x)$. Let $\alpha$ be a generator of $\mathbb{F}_{q^m}^\ast$, where $m=\textrm{ord}_{2n}(q)$. Put $\beta=\alpha^{(q^m-1)/2n}$, $m_s(x)$ is the minimal polynomial of $\beta^s, s\in X_{(n.q)}$, the reversible negacyclic code over $\mathbb{F}_q$ of length $n$ has generator polynomial $g(x)=\prod_{s\in S}\textrm{lcm}(m_s(x),m_{2n-s}(x))$, where $S\subseteq Y_{(n,q)}$.

From the BCH bound for constacyclic codes$^{[3]}$, if $C$ is a negacyclic code and let $g(x)$ be a generator polynomial of $C$ and $g(x)$ has roots $\{\beta^{1+2i}, 0\leq i\leq d-2\}$, where $\beta$  is a primitive $2n-$th root of unity, we deduce that the minimum distance of the code is at least $d$.

From the definition of negacyclic BCH codes$^{[4]}$, let $C$ be a negacyclic code with generator $g_{(q,n,\delta,b)}(x)$, then there exist an odd integer $b\geq1$ and $\delta\geq2$ such that $g(\beta^b)=g(\beta^{b+2})=\cdots=g(\beta^{b+2(\delta-2)})=0$. Then the minimum distance of the code is at least $\delta$. Denoted $C_{(q,n,\delta,b)}$ by the code with generator polynomial $g_{(q,n,\delta,b)}(x)$.\\

\dse{4~~The parameters of reversible negacyclic code} In this section, we always assume that $n=\frac{q^\ell+1}{2}$. Every negacyclic code of length $n$ is reversible from the Theorem 3.3, we will study the parameters of these codes.\\

\noindent\textbf{Lemma 4.1.} $\textrm{ord}_{2n}(q)=2\ell=m$.\\

\noindent\textbf{Proof}. It is easy to prove by Lemma 3.7.\qed\\

\noindent\textbf{Lemma 4.2.$^{[1]}$}  Let $\ell\geq2$, then any odd integer $s\in T$, $s\leq q^{\lfloor (\ell-1)/2\rfloor}+1$ and $s\not\equiv0\ \textrm{mod}q$ is a coset leader and $|T_s|=2\ell$.\\

\noindent\textbf{Theorem 4.3.} Let $n=\frac{q^\ell+1}{2}$ and $m=2\ell$. Then the minimum distance of  code $C_{(q,n,\delta,1)}, \ d\geq2\delta-1$.\\

\noindent\textbf{Proof}. Let $\alpha$ be a generator of $\mathbb{F}_{q^m}^\ast$, and put $\beta=\alpha^{q^\ell-1}$ is a primitive $2n-$th root of unity.  The generator polynomial $g_{(q,n,\delta,1)}(x)$ of code $C_{(q,n,\delta,1)}$ has roots $\beta^i, i\in \{1,3,\cdots,1+2(\delta-2)\}$ . The code $C_{(q,n,\delta,1)}$ is reversible from the Theorem 3.3. Hence we deduce that the $g_{(q,n,\delta,1)}(x)$ has roots $\beta^i, i\in \{2n-1-2(\delta-2),\cdots,2n-1,1,3,\cdots,1+2(\delta-2)\}$ . By the negacyclic BCH bound, we get $d\geq2\delta-1$.\qed\\

According to this theorem we can obtain the parameters of the code in this case. \\

\noindent\textbf{Theorem 4.4.} Let $n=\frac{q^\ell+1}{2}$. For any integer $2\leq\delta\leq \frac{q^{\lfloor(\ell-1)/2\rfloor}}{2}+2$, then the negacyclic BCH code $C_{(q,n,\delta,1)}$ has parameters
\begin{eqnarray*}
\left\{ {{\begin{array}{ll}
 {[\frac{q^\ell+1}{2},\frac{q^\ell+1}{2}-2\ell(\delta-1-\lfloor\frac{2\delta-3}{2q}\rfloor),d\geq2\delta-1]}, & {\textrm{if} \ 2\delta-3=\varepsilon+2qi},\\
 {[\frac{q^\ell+1}{2},\frac{q^\ell+1}{2}-2\ell(\delta-1-\lceil\frac{2\delta-3}{2q}\rceil),d\geq2\delta-1]}, & { \ \textrm{otherwise}}. \\
\end{array} }} \right .
\end{eqnarray*}
which $\varepsilon=2k+1<q,k\in \mathbb{Z},k\geq0$ and $i\in \mathbb{Z},i\geq0$, with generator polynomial $$\prod_{0\leq a\leq \delta-2, 1+2a\not\equiv0\ \textrm{mod}q}m_{1+2a}(x)$$
\noindent\textbf{Proof}. Since $1\leq 2\delta-3\leq q^{\lfloor(\ell-1)/2\rfloor}+1$, for any integer $a, 0\leq a\leq\delta-2$ and $1+2a\not\equiv0\ \textrm{mod}q$ is the coset leader and others are not from the Lemma 4.2. Hence we deduce that the number of $0\leq a\leq\delta-2$ and $1+2a\equiv0\ \textrm{mod}q$ is equal to $\lfloor\frac{2\delta-3}{2q}\rfloor$\ \ if $2\delta-3=\varepsilon+2qi$, where $\varepsilon=2k+1<q,k\in \mathbb{Z},k\geq0$, and $i\in \mathbb{Z},i\geq0$,  otherwise $\lceil\frac{2\delta-3}{2q}\rceil$.
By the definition of negacyclic BCH code, we get that the generator polynomial of the code is $g(x)=\Pi_{0\leq a\leq \delta-2, 1+2a\not\equiv0\ \textrm{mod}q}m_{1+2a}(x)$. According to the negacyclic code theory, the dimension of the code is equal to $n-\textrm{deg}g(x)$. Again by Lemma 4.2 and  4.3, we obtain the conclusion.\qed\\

\noindent\textbf{Corollary 4.5.} From  Theorem 4.4, let $q=3$ then we  have the following table\\

\begin{tabular}{cccc}
\hline
  & code & parameters & generator polynomial\\
  \hline
$\ell\geq3$& $C_{(3,n,3,1)}$&$[\frac{3^\ell+1}{2},\frac{3^\ell+1}{2}-2\ell,d\geq5]$ & $m_1(x)$\\

$\ell\geq3$& $C_{(3,n,4,1)}$&$[\frac{3^\ell+1}{2},\frac{3^\ell+1}{2}-4\ell,d\geq7]$ & $m_1(x)m_5(x)$\\

$\ell\geq4$& $C_{(3,n,6,1)}$&$[\frac{3^\ell+1}{2},\frac{3^\ell+1}{2}-6\ell,d\geq11]$ & $m_1(x)m_5(x)m_7(x)$\\
\hline
\end{tabular}\\

\noindent\textbf{Example 4.6.} From Corollary 4.5, we deduce \\

\begin{tabular}{ccc}
  \hline
    & code & parameters \\
  \hline
  $\ell=\{3,4,5\}$ & $C_{(3,n,3,1)}$ & $[14,8,5],[41,33,d\geq5],[122,112,d\geq5]$\\

  $\ell=\{3,4\}$ & $C_{(3,n,4,1)}$ & $[14,2,d\geq7],[41,25,d\geq7]$ \\

  $\ell=\{4,5\}$& $C_{(3,n,6,1)}$ & $[41,17,d\geq11],[122,92,d\geq11]$ \\
  \hline
\end{tabular}\\

\noindent The codes in the second row of the table are optimal linear codes from Database. \\

\dse{5~~The parameters of negacyclic BCH code} Let $n=(q^m-1)/2(q-1)$ be an integer. For studying we give the following definition. Put $\delta\geq2$ be an integer $$K_{(q,n,\delta)}=\bigcup_{0\leq i\leq \delta-1}T_{1+2i}$$ and $$-K_{(q,n,\delta)}=\{2n-a|a\in K_{(q,n,\delta)}\}$$ where $T_{1+2i}=T\cap C_{1+2i}$, and $C_{1+2i}$ is the cyclotomic coset modulo $2n$.\\

\noindent\textbf{Lemma 5.1.$^{[1]}$}\ Let $\delta=\lceil\frac{q^e}{2}\rceil$, where $e=\lfloor \frac{m-1}{2}\rfloor$, then  $K_{(q,n,\delta)}\cap(-K_{(q,n,\delta)})=\emptyset$.\\

\noindent\textbf{Theorem 5.2.}\ Let $1\leq \delta\leq \frac{q^{\lfloor(m-1)/2\rfloor}+1}{2}$ be an integer. Then the dimension  of the negacyclic BCH code $C_{(q,n,2\delta+1,1-2\delta)}$ of length $n=(q^m-1)/2(q-1)$ is equal to $$k=n-2m\large\large\lceil(2\delta-1)(q-1)/2q\rceil,$$\\
and the minimum distance of code $d\geq2\delta+1$.\\

\noindent\textbf{Proof}.\ Let the negacyclic BCH code $C_{(q,n,\delta+1,1)}$ has generator polynomial $g_{(q,n,\delta+1,1)}(x)$, by Lemma 2.4, we get $$\textrm{deg}g_{(q,n,\delta+1,1)}(x)=m\large\large\lceil(2\delta-1)(q-1)/2q\rceil.$$
The generator polynomial $g_{(q,n,2\delta+1,1-2\delta)}(x)$ of the code $C_{(q,n,2\delta+1,1-2\delta)}$ is given by $$g_{(q,n,2\delta+1,1-2\delta)}(x)=\textrm{lcm}(g_{(q,n,\delta+1,1)}(x),g_{(q,n,\delta+1,1)}^\ast(x)),$$
which $g_{(q,n,\delta+1,1)}^\ast(x))$ is the reciprocal polynomial of $g_{(q,n,\delta+1,1)}(x)$.\\  Since $1\leq \delta\leq \frac{q^{\lfloor(m-1)/2\rfloor}+1}{2}$, from Lemma 5.1, then $$g_{(q,n,2\delta+1,1-2\delta)}(x)=g_{(q,n,\delta+1,1)}(x)g_{(q,n,\delta+1,1)}^\ast(x).$$
Hence the code $C_{(q,n,2\delta+1,1-2\delta)}$ has dimension $$k=n-2m\large\large\lceil(2\delta-1)(q-1)/2q\rceil.$$ And by the negacyclic BCH bound, we have $d\geq2\delta+1$.\qed\\

\noindent\textbf{Example 5.3.} From the Theorem 5.2, we obtain the following table\\

\begin{tabular}{ccc}
  \hline
    & code & parameters \\
  \hline
  $(q,m,\delta)=(3,4,2)$ & $C_{(q,n,2\delta+1,1-2\delta)}$ & $[20,12,d\geq5]$\\

  $(q,m,\delta)=(5,4,3)$ & $C_{(q,n,2\delta+1,1-2\delta)}$ & $[78,62,d\geq7]$ \\

  $(q,m,\delta)=(5,4,4)$ & $C_{(q,n,2\delta+1,1-2\delta)}$ & $[78,54,d\geq9]$\\
  \hline
\end{tabular}\\
\\

\noindent\textbf{Lemma 5.4.}\ Let $m\geq0$ be an even integer, $\delta=\frac{q^{m/2}+1}{2}$, put $\ell=(q^{m/2}-1)/(q-1)$ when $1\leq s\leq q-1$ and $s\ell$ is an odd integer, then $s\ell$ is the coset leader of $T_{s\ell}$, $T_{s\ell}=-T_{s\ell}$ and $|T_{s\ell}|=m$, what is more,
$$K_{(q,n,\delta)}\cap(-K_{(q,n,\delta)})=\bigcup_{1\leq s\leq q-1}T_{s\ell},$$ where $s\ell$ is an odd integer.\\

\noindent\textbf{Proof}.\ It follows from the definition of $T_i=T\cap C_i$ and [1] Lemma 27.\qed\\

\noindent\textbf{Lemma 5.5.}\ Let $m\geq4$ be an even integer and $n=(q^{m/2}-1)/2(q-1)$. Put $a$ is an odd integer, $q^{(m-2)/2}\leq a\leq q^{m/2}$ and $a\not\equiv0\ \textrm{mod}q$. Then\\
(1) When $q=3$, $a$ must be a coset leader.\\
(2) When $q\geq3$ is an odd integer,  $a$ is a coset leader except that $a=1+i+i\frac{q^{m/2}-q}{q-1}, i\in I$.

1) When $m=4s, s\in \mathbb{Z}^\ast, I=\large\{\frac{q+1}{2},\frac{q+1}{2}+1,\cdots,\frac{q+1}{2}+\frac{q-5}{2}\}$.

2) When $m=4s+2, s\in \mathbb{Z}^\ast$,
i). $q=4t+1,t\in \mathbb{Z}^\ast, I=\large\{\frac{q+1}{2}+1,\frac{q+1}{2}+3,\cdots,\frac{q+1}{2}+\frac{q-7}{2}\}$

 \ \ \ \ \ \ \ \ \ \ \ \ \ \ \ \ \ \ \ \ \ \ \ \ \ \ \ \ \ \ \ \ \ \ \ \ \ \ \ ii). $q=4t-1,t\in \mathbb{Z}^\ast, I=\large\{\frac{q+1}{2},\frac{q+1}{2}+2,\cdots,\frac{q+1}{2}+\frac{q-7}{2}\}$\\
what is more, $a$ is a coset leader, then
\begin{eqnarray*}
|T_a|=
\left\{ {{\begin{array}{ll}
 {\frac{m}{2}}, & {\textrm{if} \ a=\frac{q^m+1}{2}},\\
 {m}, & {\textrm{otherwise.}} \\
\end{array} }} \right .
\end{eqnarray*}\\

\noindent\textbf{Proof}.\ By the definition of $T_i=T\cap C_i$, we get the case which  $a$ is an odd integer in the Lemma $28^{[1]}$. By binomial theorem, we deduce if $q=4t+1, t\in \mathbb{Z}^\ast$,$$\frac{q^{m/2}-q}{q-1}=\frac{m}{2}-1+\binom{\frac{m}{2}}{2}4t+\cdots+4t^{(\frac{m}{2}-1)}$$
and if $q=4t-1, t\in \mathbb{Z}^\ast$, $$\frac{q^{m/2}-q}{q-1}=\frac{m}{2}-1+\binom{\frac{m}{2}}{2}(4t-2)+\cdots+(4t-2)^{(\frac{m}{2}-1)}$$
Again classify to $m$, we will obtain the desired conclusion of lemma.\qed\\

\noindent\textbf{Theorem 5.6.}\ Let $1\leq \delta\leq \frac{q^{m/2}+1}{2}$ be an integer and $m\geq4$ be an even integer. Define $\omega=\lfloor\frac{2(\delta-1)(q-1)}{q^{m/2}-1}\rfloor$. Then the negacyclic BCH code $C_{(q,n,\delta+1,1)}$ of length $n=(q^m-1)/2(q-1)$ has minimum distance $d\geq\delta$, and dimension\\
(1) if  $\omega<\lfloor\frac{q-1}{2}\rfloor$, $k=n-m\large\lceil(2\delta-1)(q-1)/2q\rceil$\\
(2) if  $\omega\geq\lfloor\frac{q-1}{2}\rfloor$, $s,t\in \mathbb{Z}^\ast$,
\begin{eqnarray*}
k=
\left\{ {{\begin{array}{ll}
 {n-m\large\large\lceil(2\delta-1)(q-1)/2q\rceil+(2\omega-q+2)\frac{m}{2}}, & {\textrm{if}\  m=4s}, \\
{n-m\large\large\lceil(2\delta-1)(q-1)/2q\rceil+(\omega-(q-1)/2)\frac{m}{2}}, & {\textrm{if}\ m=4s+2,q=4t+1,\omega\ \textrm{is odd}},\\
{n-m\large\large\lceil(2\delta-1)(q-1)/2q\rceil+(\omega-(q-3)/2)\frac{m}{2}}, & {\textrm{if}\ m=4s+2,q=4t+1,\omega\ \textrm{is even}},\\
& {\textrm{or}\ m=4s+2,q=4t-1,\omega\ \textrm{is odd}},\\
{n-m\large\large\lceil(2\delta-1)(q-1)/2q\rceil+(\omega-(q-5)/2)\frac{m}{2}}, & {\textrm{if}\ m=4s+2,q=4t-1,\omega\ \textrm{is even}}.\\
\end{array} }} \right .
\end{eqnarray*}\\

\noindent\textbf{Proof}.\ (1) If $\omega<\frac{q-1}{2}$, it means $2(\delta-1)\leq(q^{m/2}-1)/2$.  By the Lemma 2.4 and 5.5, we obtain that every odd integer $a$ with $1\leq a\leq2\delta-1$, $a\not\equiv0\ \textrm{mod}q$ is a coset leader with $|T_a|=m$, hence $k=n-m\large\large\lceil(2\delta-1)(q-1)/2q\rceil$\\
(2) If $\omega\geq\frac{q-1}{2}$, again by Lemma 2.4 and 5.5. Let $s,t\in \mathbb{Z}^\ast$, we know,
$$\mid\{\ b\mid b\ \textrm{is odd integer and noncoset leader with}\ (q^{m/2}+1)\leq b\leq2\delta-1\ \}\mid=\mid I\mid$$.

1) $m=4s$, $|I|=\omega-\frac{q+1}{2}+1=\omega-\frac{q-1}{2}$

2) $m=4s+2$,

i). $q=4t+1$, and $\omega$ is even, then $|I|=\frac{\omega-\frac{q+1}{2}-1}{2}+1=\frac{2\omega-q+1}{4}$

ii). $q=4t+1$, and $\omega$ is odd, then $|I|=\frac{\omega-\frac{q+1}{2}-1-1}{2}+1=\frac{2\omega-q-1}{4}$

iii). $q=4t-1$, and $\omega$ is even, then $|I|=\frac{\omega-\frac{q+1}{2}}{2}+1=\frac{2\omega-q+3}{4}$

iv). $q=4t-1$, and $\omega$ is odd, then $|I|=\frac{\omega-\frac{q+1}{2}-1}{2}+1=\frac{2\omega-q+1}{4}$\\
Again by Lemma 2.4 and 5.5, when $a$ is odd with $1\leq a\leq2\delta-1$, the number of coset leader is equal to $\lceil(2\delta-1)(q-1)/2q\rceil-|I|$. And put $\widetilde{a}=(q^{m/2}+1)/2\in T$, $|T_{\widetilde{a}}|=\frac{m}{2}$, then the proof of the theorem is completed.\qed\\

\noindent\textbf{Theorem 5.7.}\ Let $\delta=\frac{q^{m/2}+1}{2}$ be an integer and $m\geq4$ be an even integer, then the negacyclic BCH code $C_{(q,n,\delta,1)}$ with length $n=(q^m-1)/2(q-1)$ has minimum distance $d\geq\delta$ and dimension\\
\begin{eqnarray*}
k=
\left\{ {{\begin{array}{ll}
{n-\large\frac{m(q-1)}{2}q^{(m-2)/2}+\frac{qm}{2}}, & {\textrm{if}\ m=4s},\\
{n-\large\frac{m(q-1)}{2}q^{(m-2)/2}+\frac{(q+1)m}{4}}, & {\textrm{if}\ m=4s+2,q=4t+1},\\
{n-\large\frac{m(q-1)}{2}q^{(m-2)/2}+\frac{(q+3)m}{4}}, & {\textrm{if}\ m=4s+2,q=4t-1}.\\
\end{array} }} \right .
\end{eqnarray*}\\where $s,t\in \mathbb{Z}^\ast$\\

\noindent\textbf{Proof}.\ In this case, $\lceil(2\delta-1)(q-1)/2q\rceil=\frac{(q-1)}{2}q^{(m-2)/2}$, $\omega=q-1>\frac{q-1}{2}$ is a even integer, the proof of dimension follows from Theorem 5.6, and by the negacyclic BCH bound , we have minimum distance $d\geq\delta$.\qed\\

\noindent\textbf{Theorem 5.8.}\ Let $1\leq \delta\leq \frac{q^{m/2}+1}{2}$ be an integer and $m\geq4$ be an even integer. Define $\omega=\lfloor\frac{2(\delta-1)(q-1)}{q^{m/2}-1}\rfloor, \varpi=\lfloor\frac{(2\delta-1)(q-1)}{q^{m/2}-1}\rfloor$. Then the reversible negacyclic BCH code $C_{(q,n,2\delta+1,1-2\delta)}$ of length $n=(q^m-1)/2(q-1)$ has minimum distance $d\geq2\delta+1$ and dimension\\
(1) if  $\omega<\lfloor\frac{q-1}{2}\rfloor$,
\begin{eqnarray*}
k=
\left\{ {{\begin{array}{ll}
 {n-m\large\lceil(2\delta-1)(q-1)/2q\rceil}, & {\textrm{if}\ m=4s},\\
 {n-m\large\lceil(2\delta-1)(q-1)/2q\rceil+\lceil \varpi/2\rceil m}, & {\textrm{if}\  m=4s+2}. \\
\end{array} }} \right .
\end{eqnarray*}\\
(2) if  $\omega\geq\lfloor\frac{q-1}{2}\rfloor$, $s,t\in \mathbb{Z}^\ast$,
\begin{eqnarray*}
k=
\left\{ {{\begin{array}{ll}
 {n-2m\large\large\lceil(2\delta-1)(q-1)/2q\rceil+(2\omega-q+2)m}, & {\textrm{if}\  m=4s}, \\
{n-2m\large\large\lceil(2\delta-1)(q-1)/2q\rceil+(\omega-(q-1)/2+\lceil \varpi/2\rceil)m}, & {\textrm{if}\ m=4s+2,q=4t+1,\omega\ \textrm{is odd}},\\
{n-2m\large\large\lceil(2\delta-1)(q-1)/2q\rceil+(\omega-(q-3)/2+\lceil \varpi/2\rceil)m}, & {\textrm{if}\ m=4s+2,q=4t+1,\omega\ \textrm{is even}},\\
& {\textrm{or}\ m=4s+2,q=4t-1,\omega\ \textrm{is odd}},\\
{n-2m\large\large\lceil(2\delta-1)(q-1)/2q\rceil+(\omega-(q-5)/2+\lceil \varpi/2\rceil)m}, & {\textrm{if}\ m=4s+2,q=4t-1,\omega\ \textrm{is even}}.\\
\end{array} }} \right .
\end{eqnarray*}\\

\noindent\textbf{Proof}.\ Let the negacyclic BCH code $C_{(q,n,\delta+1,1)}$ has generator polynomial $g_{(q,n,\delta+1,1)}(x)$, then we have \\
1) if $\omega<\frac{q-1}{2}$, deg$g_{(q,n,\delta+1,1)}(x)=m\large\lceil(2\delta-1)(q-1)/2q\rceil$\\
2) if $\omega\geq\frac{q-1}{2}$ and $s,t\in \mathbb{Z}^\ast$,
\begin{eqnarray*}
\textrm{deg}g_{(q,n,\delta+1,1)}(x)=
\left\{ {{\begin{array}{ll}
{m\lceil(2\delta-1)(q-1)/2q\rceil+(2\omega-q+2)\frac{m}{2}}, & {\textrm{if}\  m=4s}, \\
{m\lceil(2\delta-1)(q-1)/2q\rceil+(\omega-(q-1)/2)\frac{m}{2}}, & {\textrm{if}\ m=4s+2,q=4t+1,\omega\ \textrm{is odd}},\\
{m\lceil(2\delta-1)(q-1)/2q\rceil+(\omega-(q-3)/2)\frac{m}{2}}, & {\textrm{if}\ m=4s+2,q=4t+1,\omega\ \textrm{is even}},\\
& {\textrm{or}\ m=4s+2,q=4t-1,\omega\ \textrm{is odd}},\\
{m\lceil(2\delta-1)(q-1)/2q\rceil+(\omega-(q-5)/2)\frac{m}{2}}, & {\textrm{if}\ m=4s+2,q=4t-1,\omega\ \textrm{is even}}.\\
\end{array} }} \right .
\end{eqnarray*}\\
From the definition of $C_{(q,n,2\delta+1,1-2\delta)}$, which has generator polynomial $g_{(q,n,2\delta+1,1-2\delta)}(x)$,$$g_{(q,n,2\delta+1,1-2\delta)}(x)=\textrm{lcm}(g_{(q,n,\delta+1,1)}(x),g_{(q,n,\delta+1,1)}^\ast(x))$$
where $g_{(q,n,\delta+1,1)}^\ast(x)$ is the reciprocal polynomial of $g_{(q,n,\delta+1,1)}(x)$.\\
It follows from Lemma 5.4, we deduce,
\begin{eqnarray*}
\textrm{deg}(\textrm{gcd}(g_{(q,n,\delta+1,1)}(x),g_{(q,n,\delta+1,1)}^\ast(x)))=
\left\{ {{\begin{array}{ll}
 {0}, & {\textrm{if}\ m=4s},\\
 {\lceil\varpi/2\rceil}, & {\textrm{if}\  m=4s+2}. \\
\end{array} }} \right .
\end{eqnarray*}\\
Hence, $$\textrm{deg}g_{(q,n,2\delta+1,1-2\delta)}(x)=2\textrm{deg}g_{(q,n,\delta+1,1)}(x)-\textrm{deg}(\textrm{gcd}(g_{(q,n,\delta+1,1)}(x),g_{(q,n,\delta+1,1)}^\ast(x))).$$
Then the dimension can be obtained, and by the negacyclic BCH bound, we have minimum distance $d\geq2\delta+1$.\qed\\

\noindent\textbf{Theorem 5.9.}\ Let $\delta=\frac{q^{m/2}+1}{2}$ be an integer and $m\geq4$ be an even integer, then the negacyclic BCH code $C_{(q,n,2\delta+1,1-2\delta)}$ with length $n=(q^m-1)/2(q-1)$ has minimum distance $d\geq2\delta+1$ and dimension\\
\begin{eqnarray*}
k=
\left\{ {{\begin{array}{ll}
{n-(q-1)q^{(m-2)/2}m+(q+1)m}, & {\textrm{if}\ m=4s+2, q=4t-1},\\
{n-(q-1)q^{(m-2)/2}m+qm}, & {\textrm{otherwise}}.\\
\end{array} }} \right .
\end{eqnarray*}\\Where $s, t\in \mathbb{Z}^\ast$\\

\noindent\textbf{Proof}.\ In this case, $\lceil(2\delta-1)(q-1)/2q\rceil=\frac{(q-1)}{2}q^{(m-2)/2}$, $\omega=q-1>\frac{q-1}{2}$ is an even integer, and $\varpi=q-1$. The dimension follows from Theorem 5.8, and by the negacyclic BCH bound, we have minimum distance $d\geq2\delta+1$.\qed\\

\noindent\textbf{Example 5.10.}\ From the Theorem 5.9, we have following table\\

\begin{tabular}{ccc}
  \hline
    & code & parameters \\
  \hline
  $(q,m,\delta)=(3,6,14)$ & $C_{(q,n,2\delta+1,1-2\delta)}$ & $[182,98,d\geq29]$\\

  $(q,m,\delta)=(5,4,13)$ & $C_{(q,n,2\delta+1,1-2\delta)}$ & $[78,18,d\geq27]$ \\
  \hline
\end{tabular}\\
\\

Let $n=\frac{q^{t\cdot2^\tau}-1}{2(q^t+1)}$ be an integer, where $t,\tau\in \mathbb{Z}^\ast, \tau\geq2, t\geq2$. We study those negacyclic BCH codes of length $n$.\\

\noindent\textbf{Lemma 5.11.} Let $m=\textrm{ord}_{2n}(q)$, then $m=t\cdot2^\tau$.\\

\noindent\textbf{Proof}. Since $2n\mid (q^{t\cdot2^\tau}-1)$, then $m\mid t\cdot2^\tau$. We get $m\in\{2^i,\ t,\ t\cdot2^i \mid 0\leq i\leq\tau\}$. Write $2n$ to the base $q$ as $2n=\sum_{j=0}^{s_{2n}}jq^j$, where $0\leq j\leq q-1$. As $2n=\frac{q^{t\cdot2^\tau}-1}{q^t+1}\geq\frac{q^{t\cdot2^\tau}-1}{2q^t}$, then $s_{2n} \geq (t\cdot2^\tau-t-1)$. And since $2n\mid (q^m-1)$, hence $m\geq(t\cdot2^\tau-t-1)> t\cdot2^{\tau-1}$. Thus $m=t\cdot2^\tau$. \qed\\

\noindent\textbf{Lemma 5.12.}  For any odd integer $s\in T$, $s\leq q^{\lfloor (t\cdot2^{\tau-1}-1)/2\rfloor}+1$ and $s\not\equiv0\ \textrm{mod}q$ is a coset leader and $|T_s|=t\cdot2^\tau$.\\

\noindent\textbf{Proof}. It follows from the Lemma 4.2. \qed\\

\noindent\textbf{Theorem 5.13.} Let $n=\frac{q^{t\cdot2^\tau}-1}{2(q^t+1)}$, where $t,\tau\in \mathbb{Z}^\ast, \tau\geq2, t\geq2$. For any integer $\delta$, $1\leq\delta\leq \frac{q^{\lfloor (t\cdot2^{\tau-1}-1)/2\rfloor+1}}{2}$, then the dimension of negacyclic BCH code $C_{(q,n,2\delta+1, 1-\delta)}$, $$k=n-t\cdot2^{\tau+1}\large\large\lceil(2\delta-1)(q-1)/2q\rceil,$$
and $d\geq2\delta+1$.\\

\noindent\textbf{Proof}. From Theorem 5.2, Lemma 5.11 and 5.12, we can obtain the conclusion immediately.\qed\\

\noindent\textbf{Example 5.14.}\ We get the following table \\

\begin{tabular}{ccc}
  \hline
    & code & parameters \\
  \hline
  $(q,t,\tau,\delta)=(3,2,2,2)$ & $C_{(q,n,2\delta+1,1-2\delta)}$ & $[328,312,d\geq5]$\\
  $(q,t,\tau,\delta)=(3,2,2,3)$ & $C_{(q,n,2\delta+1,1-2\delta)}$ & $[328,296,d\geq7]$\\
  \hline
\end{tabular}\\

\dse{6~~A class of MDS LCD negacyclic codes} In this section, we study the MDS LCD negacyclic codes over $\mathbb{F}_q$. A $[n,k,d]$ code is called maximum distance separable (abbreviated MDS) if $d=n-k+1$. Let $\beta$ be a primitive $2n-$th root of unity in $\mathbb{F}_{q^m}$, where $m=\textrm{ord}_{2n}(q)$. Let $C$ be a negacyclic code over $F_q$ of length $n$. Let $S$ be the defining set of $C$. Then $S$ is the union of some $T_i$, where $T_i$ is defined above. Then we have following theorem.\\

\noindent\textbf{Theorem 6.1.} Let $n$ be an even integer and $n\mid(q-1)$. Let $C$ be the negacyclic code over $F_q$ of length $n$ with defining set $S\subset T$. Put $0\leq\rho<\frac{n}{2}-1$. Then the $C$ is a MDS LCD negacyclic code if $$S=\{1+2i\mid i=\frac{n}{2}+j\ \textrm{and}\ \frac{n}{2}-k, 0\leq j\leq\rho,\ 1\leq k\leq \rho+1\}\ (\textrm{mod}2n)$$ moreover $C$ has parameters $[n,n-2(\rho+1),2\rho+3]$\\

\noindent\textbf{Proof}. We can easily get that $S$ has $2(\rho+1)$ elements, and then the dimension of $C$ is $n-2(\rho+1)$. Put $A=\{i\mid i=\frac{n}{2}+j\ \textrm{and}\ \frac{n}{2}-k, 0\leq j\leq\rho,\ 1\leq k\leq \rho+1\}$. Then $A$ has $2(\rho+1)$ consecutive integers. From the negacyclic BCH bound and the singleton bound, we can get that the minimum distance of $C$ is $2\rho+3$, and the code $C$ is a MDS code.

Next we illustrate that the code $C$ is a LCD code. $\forall\ a\in S$, there exist two integer $j_0$,$k_0$. $0\leq j_0\leq\rho$ or $1\leq k_0\leq\rho+1$ such that $a=n+2j_0+1$ or $a=n-2k_0+1$. If $a=n+2j_0+1$, then $2n-a=1+2(\frac{n}{2}-(j_0+1))$, and $1\leq j_0+1\leq(\rho+1)$. Hence, $2n-a\in S$. If $a=n-2k_0+1$, then $2n-a=1+2(\frac{n}{2}+(k_0-1))$, and $0\leq k_0-1\leq\rho$. Hence, $2n-a\in S$.  Summarizing the two cases get $\forall\ a\in S$, then $2n-a\in S$. By Lemma 3.4, we deduce that $C$ is a LCD code. \qed\\

\noindent\textbf{Example 6.2.} We list some MDS LCD negacyclic codes over some finite fields in the following table.\\

\begin{tabular}{cccc|cccc|cccc}
  \hline
   $q$ & $n$ & $\rho$ & parameters & $q$ & $n$ & $\rho$ & parameters & $q$ & $n$ & $\rho$ & parameters\\
  \hline
  5 & 4 & 0 & $[4,2,3]$ &  11 & 10 & 3 & $[10,2,9]$ & 17 & 8 &1 & $[8,4,5]$\\
  7 & 6 & 0 & $[6,4,3]$ &  13 & 6 & 0 & $[6,4,3]$ & 17 & 8 & 2 & $[8,2,7]$\\
  7 & 6 & 1 & $[6,2,5]$ &  13 & 6 & 1 & $[6,2,5]$ & 17 & 16 & 0 & $[16,14,3]$\\
  9 & 4 & 0 & $[4,2,3]$ &  13 & 12 & 0 & $[12,10,3]$ & 17 & 16 & 1 & $[16,12,5]$\\
  9 & 8 & 0 & $[8,6,3]$ &  13 & 12 & 1 & $[12,8,5]$ & 17 & 16 & 2 & $[16,10,7]$\\
  9 & 8 & 1 & $[8,4,5]$ &  13 & 12 & 2 & $[12,6,7]$ & 17 & 16 & 3 & $[16,8,9]$\\
  9 & 8 & 2 & $[8,2,7]$ &  13 & 12 & 3 & $[12,4,9]$ & 17 & 16 & 4 & $[16,6,11]$\\
  11 & 10 & 0 & $[10,8,3]$ & 13 & 12 & 4 & $[12,2,11]$ & 17 & 16 & 5 & $[16,4,13]$\\
  11 & 10 & 1 & $[10,6,5]$ & 17 & 4 & 0 & $[4,2,3]$ & 17 & 16 & 6 & $[16,2,15]$\\
  11 & 10 & 2 & $[10,4,7]$ & 17 & 8 & 0 & $[8,6,3]$\\
  \hline
\end{tabular}\\

\dse{7~~Conclusion}We have studied the reversible negacyclic codes over finite fields. In section 2, we deduce the condition of reversible negacycylic codes. In section 3, the structure of LCD negacyclic codes is determined, and in the special case, the quantity of reversible negacyclic codes is gained. In section 4 and 5, we discuss the parameters of negacyclic BCH codes when the length $n=\frac{q^\ell+1}{2}$ , $n=\frac{q^m-1}{2(q-1)}$ and $n=\frac{q^{t\cdot2^\tau}-1}{2(q^t+1)}$. In section 6, we study a class of MDS LCD negacyclic codes.
\\
\\
\\
\\


\begin{thebibliography}{99}
\bibitem{pa} C. S. Ding, C. J. Li and S. X. Li, LCD cyclic Codes over Finite Fields, arXiv:1608.02170v1[cs.IT].

\bibitem{pa} A. Krishna, Dilip V.Sarwate,  Pseudocyclic Maximum-Distance Separable codes, IEEE Trans, Inf. Theory, vol.36, No.4(1990), pp.880-884.

\bibitem{pa} N. Aydin, I. Siap, D. K. Ray-Chaudhuri, The structure of 1-generator quasi-twisted codes and new linear codes, Des. Codes Cryptogr. 24 (2001), 313-326.

\bibitem{pa} G. G. La Guardia, On negacyclic MDS-convolutional codes, Linear Algebra and its Applications, 448 (2014), 85-96.

\bibitem{pa} C. J. Li, C. S. Ding and H. Liu, Parameters of two classes of LCD BCH codes, arXiv:1608.02670v1[cs.IT].

\bibitem{pa}  J. L. Massey, Linear codes with complementary duals, Discrete Math, vol. 106/107(1992), pp. 337-342.

\bibitem{pa} X. Yang and J. L. Massey, The condition for a cyclic code to have a complementary  dual, Discrete Math, vol. 126(1994), pp.391-393.

\bibitem{pa} N. Sendrier, Linear codes with complementary duals meet the Gilbert-Varshamov bound, Discrete Math, vol. 285(2004), pp.345-347.
\bibitem{pa} X. Hou, and F. Oggier, On LCD codes and lattices,  IEEE International Symposium on Information Theory, 2016, pp.1501-1505.

\bibitem{pa} E. R. Lina Jr. and  E. G. Nocon, On the construction of some LCD codes over finite fields,  the DLSU Research Congress, vol.4(2016).

\bibitem{pa} H. Q. Dinh and  S. R. L\'{o}pez-Permouth, Cyclic and negacyclic codes over finite chain rings, IEEE Trans, Inf. Theory, vol.50, No.8(2004), pp.1728-1744.
\bibitem{pa} Y. S. Yang and  W. C. Cai, On self-dual constacyclic codes over finite fields, Des. Codes Cryptogr, vol.74(2015), pp.335-364.
\bibitem{pa} C. G\"{u}neri, B. \"{O}zkaya and P. Sol\'{e}, Quasi-cyclic complementary dual codes, Finite Fields and Their Applications, vol.42 (2016), pp.67-80.
\end{thebibliography}
\end{document}